\documentclass[12pt]{iopart}
\usepackage{iopams}

\usepackage[figuresright]{rotating}

\begin{document}

\title{TeV gamma-rays from the Northern sky pulsar wind nebulae}

\author{W. Bednarek \& M. Bartosik}

\address{Department of Experimental Physics, University of \L \'od\'z,
ul. Pomorska 149/153, 90-236 \L \'od\'z, Poland\\

E-mail:  bednar@fizwe4.fic.uni.lodz.pl; bartosik@uni.lodz.pl}

\begin{abstract}
We estimate the TeV $\gamma$-ray fluxes expected from the 
population of young pulsars in terms of the 
self-consistent time dependent hadronic-leptonic model for 
the high energy processes inside the pulsar wind nebulae (PWNe).
This radiation model bases on the hypothesis of Arons and collaborators
who postulate that leptons are accelerated inside the nebulae 
as a result of resonant scattering on heavy nuclei, which in turn are 
accelerated in 
the pulsar wind region or the pulsar inner magnetosphere. Our aim is to find out 
which PWNe on the northern hemisphere are the best candidates for detection 
at energies above 60 GeV and 200 GeV by 
the next generation of low threshold Cherenkov telescopes. 
\end{abstract}


\submitto{\JPG}



%
%
\section{Introduction}

Young pulsars lose their rotational energy in the 
form of relativistic winds in which particles can be accelerated to very high energies.  
These winds interact with the surrounding supernova, creating the pulsar wind nebulae
observed from radio to X-ray energies. A few tens of such objects have 
been observed in TeV $\gamma$-rays (e.g. Srinivasan et al.~1997, 
Chadwick et al.~1999, Aharonian et al.~2001, Aharonian et al.~2002), 
but only four of them were detected in the past, i.e. 
the Crab Nebula (Weekes et al.~1989), the nebulae around the Vela pulsar
(Yoshikoshi et al.~1997), PSR 1706-44 (Kifune et al.~1995, Chadwick et al.~1998), 
and PSR 1509-58 (Sako et al.~2000, Aharonian et al. 2005a). 
However, recent observation of the PSR 1706-44 by 
the HESS group (Aharonian et al.~2005b) does not confirm the detection of the TeV 
emission from this nebula. 
The estimated upper limit, on the level of a few percent of the Crab Nebula 
flux, is significantly below the flux reported previously.
The observation of the $\sim 80$ TeV $\gamma$-rays from the Crab Nebula 
(Horns et al.~2003, Aharonian et al. 2004) indicate the existence of particles with 
energies up to $\sim 10^{15}$ eV at least inside this PWNa. Therefore, PWNe 
seem to be
one of the best candidates for the sites of cosmic ray acceleration
(e.g. Bednarek \& Protheroe~2002, Giller \& Lipski~2002, Bednarek \& Bartosik~2004).     
However, details of the acceleration process inside the PWNe are mainly unknown.    
There is a hope that observation and modelling of the multiwavelength emission from 
such objects can provide an important information allowing to solve this problem.
It is widely argued that the radiation in the PWNe is produced by leptons 
in the synchrotron and the inverse Compton processes (e.g. de Jager \& Harding~1992, 
Atoyan \& Aharonian~1996, Hillas et al.~1998, Aharonian et al. 2004). Possible 
contribution of hadrons at the highest energies has been also considered 
(e.g. Atoyan \& Aharonian~1996, Bednarek \& Protheroe~1997, Bednarek \& Bartosik~2003). 

In this paper we consider the time dependent hadronic-leptonic model for the 
production of radiation inside the PWNe basing on the acceleration model of
hadrons and leptons inside the nebula, proposed by Arons and collaborators
(see e.g. Galant \& Arons~1994, Arons~1998).
In this model we calculate self-consistently the multiwavelength spectrum produced by 
leptons and the high energy radiation generated by hadrons also responsible for 
the acceleration of leptons.
The model is applied to the PWNe which have not been observed in 
the TeV $\gamma$-rays yet. We predict the expected fluxes from the population of
relatively young PWNe on the northern hemisphere (with ages lower than 
$\sim 10^5$ yrs). The earlier modelling of the high 
energy processes in the PWNe, which base on pure leptonic models (e.g. 
Aharonian et al.~1997), give a list of pulsars on the 
whole sky which are the best candidates for TeV $\gamma$-ray observations due to their 
highest ratio of the spin down luminosity divided by the square of distance.
Here we consider young pulsars which can be observed at the Northern 
hemisphere.
Some of them are well known (and already considered by Aharonian et al. 1997) but 
others have been recently reported in the Parces Multibeam pulsar survey catalogs 
(Morris et al.~2002, Kramer et al.~2003) and  
original papers (e.g. Torii et al.~1997, Halpern et al.~2001, Camilo et al.~2002,
Roberts et al.~2002, McLaughlin et al.~2002).

\section{Acceleration of particles inside the PWN}

Pulsars with short periods and strong surface magnetic fields, created
during the supernova explosions, are surrounded by the non-thermal compact nebulae
(the so called pulsar wind nebulae - PWNe) observed from radio up to X-rays. 
These non-thermal nebulae are immersed in the expanding supernova remnants which shells 
are also observed in some cases from the radio up to X-rays. 
The PWNe contain relativistic
leptons responsible for non-thermal emission from radio up to TeV $\gamma$-rays. 
The larger scale supernova remnants contain most of the mass of the expanding 
supernova. If hadrons are accelerated in the inner part of the PWNa, they can diffuse 
to the supernova remnant and interact with the mass of the supernova producing
additional $\gamma$-rays.

Here we consider a scenario in which rotating magnetospheres of neutron stars can
accelerate not only leptons but also heavy nuclei, extracted from
positively charged polar cap regions. In fact, different aspects of the high energy
phenomena around pulsars, such as the change in the drift direction of the radio
subpulses (Gil et al.~2003), the existence of morphological features inside the Crab 
Nebula, and the appearance of extremely energetic leptons inside it 
(Gallant \& Arons 1994), can be naturally explained by the presence of heavy nuclei.
Arons and collaborators
(e.g. see Arons~1998) postulate that the Lorentz factors of iron nuclei, 
accelerated somewhere in the inner magnetosphere and/or the Crab pulsar 
wind zone should be,
\begin{eqnarray}
\gamma_{Fe}\approx \eta Ze\Phi_{\rm open}/m_{\rm Fe}c^2\approx
8\times 10^9 \eta B_{12} P_{\rm ms}^{-2},
\label{eq1}
\end{eqnarray}
\noindent
where $m_{\rm Fe}$ and $Ze$ are the mass and charge of the iron nuclei,
$c$ is the velocity of light, and $\Phi_{\rm open} = \sqrt{L_{\rm rot}/c}$
is the total electric potential drop across the open magnetosphere,
$L_{\rm rot}(t) = B_{\rm s}^2 R_{\rm s}^6 \Omega^4/6c^3\approx 
3\times 10^{43}B_{12}^2P_{\rm ms}^{-4}~~{\rm erg~s}^{-1}$, 
\noindent
is the pulsar energy loss rate for the emission of the magnetic dipole radiation,
$\Omega = 2\pi/P$, and the period of the pulsar, $P = 10^{-3}P_{\rm ms}$ s,
changes with time according to 
$P^2_{\rm ms}(t) = P_{\rm 0,ms}^2 + 2\times 10^{-9}B_{12}^2t$, 
\noindent
where $P_{\rm 0,ms}$ is the initial period of the pulsar, $B = 10^{12}B_{12}$ G is 
the surface magnetic field of the pulsar,
and $\eta$ is the acceleration factor which determines the Lorentz factor of 
nuclei in respect to the maximum allowed by the pulsar electrodynamics.
Following Arons and collaborators, 
we assume that: (1) $\eta$ is not very far from unity, adopting the value $\eta = 0.5$; 
(2) iron nuclei take most of the spin down power of the pulsar, 
$L_{\rm Fe} = \chi L_{\rm rot}$, where $\chi = 0.95$. 
Unfortunately, this values are not predicted at present by 
any model for the acceleration of ions in the pulsar wind and can only be constrained  
by the high energy observations of the PWNe.
The iron nuclei can be extracted from the neutron star surface and accelerated
during the pulsar radio phase when the efficient leptonic cascades heat
the polar cup region. For more details of the acceleration and propagation of nuclei
inside the pulsar magnetosphere we refer to Bednarek \& Bartosik (2003).
The iron nuclei generate Alfven waves in the down-stream region of the pulsar
wind shock, which energy is resonantly transfered to leptons
present in the wind (Hoshino et al.~1992). 
As a result, leptons obtain a spectrum close to a power law with the
spectral index $\delta_1\approx 2$ between $E_{\rm 1} = \gamma_{\rm Fe}m_{\rm e}c^2$ 
and $E_{\rm 2}\approx \gamma_{\rm Fe} A m_{\rm p}c^2/Z$ (see Gallant \& Arons~1994), 
where $m_{\rm e}$ and $m_{\rm p}$ are the electron and proton mass, respectively.
The spectrum is normalized to the conversion efficiency of energy from the iron nuclei 
to the positrons, $\xi$. Note that the radiation from  
leptons depends on the product of the energy conversion from the pulsar wind
to nuclei, $\chi$, and the acceleration efficiency of positrons by these ions, $\xi$.
Therefore, decreasing the first coefficient and increasing the second one
obtains the same level of radiation from positrons but a lower level of gamma-ray
flux from hadronic interactions of ions with the matter inside the nebula.
Since the dependence of $\chi \cdot \xi$ on time is not predicted by any 
theoretical model we keep this value constant during the evolution of the nebula. 
Relativistic particles accelerated by the mechanism discussed above are captured 
inside the pulsar wind nebula losing energy on different processes. 

\section{Production of gamma-rays}

Nuclei and leptons are injected into the pulsar wind nebula which parameters 
are changing significantly during its evolution. To take into account this effect 
we have constructed a simple model for the time evolution of the nebula under 
the influence of the pulsar following the previous works (e.g. Ostriker \& Gunn~1971, 
Pacini \& Salvati~1973). 
 The basic parameters of the expanding nebula as a function of time such as:
its outer radius and the radius of the pulsar wind shock, the velo\-ci\-ty of 
expansion, the mass of the nebula, and the strength of the magnetic field 
inside the nebula, are determined by applying the step time numerical method.
Numerical approach allows us to take into account the effects of energy
conversion from nuclei to the expanding nebula due to their adiabatic energy
losses. The model for the expansion of the nebula has to be relatively simple at this
stage of calculations in order to take into account properly the time dependent radiation 
processes inside the nebula at different times after supernova explosion. 
Having in hand such a model, we can calculate the 
equilibrium spectra of leptons and nuclei at an arbitrary age of the nebula taking into 
account different energy loss processes. 
Leptons injected into the medium of the expanding supernova 
remnant suffer energy losses mainly on radiation processes, bremsstrahlung, synchrotron, 
and the inverse Compton, and due to the expansion of the nebula. 
The rate of their energy losses can be described by 
\begin{eqnarray}
-{{dE}\over{dt}} = (\alpha_1 + \alpha_2)E + (\beta_1 + \beta_2) E^2~~~{\rm GeV~s^{-1}},
\label{eq22}
\end{eqnarray}
\noindent
where $\alpha_1\approx 7.8\times 10^{-16} N$ s$^{-1}$ 
describes the bremsstrahlung losses, where $N$ is the number density of the medium in
cm$^{-3}$; $\alpha_2 = V_{\rm Neb}(t)/R_{\rm Neb}(t)$ describes the adiabatic losses
due to the expansion of the nebula (Longair 1981) where $V_{\rm Neb}(t)$ and 
$R_{\rm Neb}(t)$ are the velocity of expansion and the  radius of the nebula at 
the time t; $\beta_1\approx 2.55\times 
10^{-6}B^2$ GeV$^{-1}$ s$^{-1}$ the synchrotron energy losses, where $B$ is 
the magnetic field in G; and $\beta_2\approx 1.05\times 
10^{-7}U_{\rm rad}$ GeV$^{-1}$ s$^{-1}$, $U_{\rm rad}$ is the energy density of different 
types of radiation inside the nebula in GeV cm$^{-3}$, the ICS losses in the Thomson 
regime in different types of soft radiation inside the 
nebula, i.e. the synchrotron radiation produced by leptons in the magnetic field of the 
nebula, the microwave background radiation (MBR), and the infrared photons 
emitted by dust inside the nebula. 
The energy losses of leptons on the ICS in the Klein-Nishina regime 
can be safely neglected in respect to the synchrotron energy losses. The density of
synchrotron radiation depends on the spectrum of leptons which is in turn determined 
by their energy losses at the earlier phase of expansion of the nebula. 

The coefficients, $\alpha_1, \alpha_2, \beta_1$, and $\beta_2$, depend on time 
in a complicated way due to the changing conditions in the expanding nebula (magnetic
field, density of matter and radiation). Therefore, Eq.~\ref{eq22} can not be integrated 
analytically for the arbitrary time after supernova explosion. 
In order to determine the energies of leptons, $E$, inside the 
nebula at a specific time $t_{\rm obs}$
which have been injected with energies $E_{\rm o}$ at an earlier time $t$ 
we use the numerical approach. The step time method is applied in which the 
conditions in the nebula at the relatively short period, $t$ to $(t+\Delta t)$, 
are assumed constant.
Applying the parameters of the nebula determined for the time $t$  
the energies of leptons after the next time step $\Delta t$ are determined 
by solving equation Eq.~\ref{eq22} analytically. 
Next, the conditions inside the
nebula are changed to values which are obtained from the expansion model of the nebula
at the time $t+\Delta t$.
The equilibrium spectrum of leptons at the time, $t_{\rm obs}$, is then obtained  
by summing over the spectra injected at specific time and over all time steps
up to the present observed time $t_{\rm obs}$, 
\begin{eqnarray}
{{dN(t_{\rm obs})}\over{dE}} = \sum_{t=0}^{t=t_{\rm obs}} J(t')
 {{dN}\over{dE_{\rm o}}dt}dt,
\label{eq24}
\end{eqnarray}
\noindent
where $dN/dE_{\rm o}dt$ is the injection spectrum of leptons at the time t, 
$t' = t_{\rm obs} - t$, and the jacobian $J(t') = E_{\rm o}/E(t')$ which
describes the change of energy of lepton during the period $\Delta t$
is calculated analitycally by solving Eq.~2.
The example spectra of leptons inside the nebula at the specific time after explosion
of supernova and more details of these calculations are given 
in Bednarek \& Bartosik~(2003). 

The knowledge on the equilibrium spectra of relativistic leptons 
as a function of time after supernova explosion allows us to calculate the photon 
spectra produced inside the nebula by these particles in different radiation processes. 
Leptons produce photons mainly in synchrotron, bremsstrahlung, and ICS processes. 
For the $\gamma$-ray energies, which are of interest in this paper (above 60 GeV and 
200 GeV), the contribution of $\gamma$-rays from interaction of hadrons with the matter 
is negligible (see Bednarek \& Bartosik~2003). Therefore hadronic $\gamma$-rays
are not considered in 
this paper (they are calculated in Bednarek \& Bartosik~(2003) where the 
$\gamma$-ray spectra in the whole energy region are presented). 

The conditions in the expanding nebula, i.e. the magnetic and radiation fields and the
density of matter, change significantly with time in a different manner. Thus, the 
relative importance of specific radiation processes has to change as well. At the early 
stage of expansion of the nebula, the synchrotron energy losses of leptons dominate over 
the ICS and the bremsstrahlung energy losses. Therefore, most of the energy of leptons is 
radiated in the low energy range. When the nebula becomes older, the energy density 
of the synchrotron radiation inside the nebula decreases but the energy density of 
the MBR remains constant. Relative
importance of the ICS losses increases with respect to the synchrotron energy losses.
In fact, for the PWNe with the age $\ge 10^4$ yrs, the $\gamma$-rays are produced
mainly by leptons scattering the MBR (not synchrotron photons). 
Therefore, the possible displacement of the pulsar from the place of its origin
has no effect on the expected level of the $\gamma$-ray flux but only on the dimensions
of the $\gamma$-ray source. 

In order to determine precise contributions of these three radiation processes, 
we calculate the photon spectra produced by leptons with the equilibrium spectrum
calculated for different times after supernova explosion.
The multiwavelength spectra emitted by the pulsar wind nebulae with specific 
parameters at different times after supernova explosion are shown in Figs. 4 and 5 in 
Bednarek \& Bartosik~(2003). 

\section{Gamma-ray fluxes from specific PWNe}

The model described above has been confronted with the known PWNe which have been
reported as 
sources of high energy $\gamma$-rays. In order to calculate the multiwavelength 
spectra from specific PWNe, we have to fix some initial parameters of their pulsars 
and the expanding supernova envelopes. We apply the masses and the initial 
expansion velocities for most of the considered nebulae as observed in the case of 
the Crab Nebula, i.e. equal to $4.6\pm 1.8$ M$_\odot$ (Fesen et al.~1997) and 2000 
km s$^{-1}$ (Davidson \& Fesen~1985). We are aware that the Crab Nebula is not 
typical in many ways but our aim is to
give an order of estimate of the TeV fluxes for the future observations with the 
next generation Cherenkov telescopes.
In the case of nebulae for which the ages can be estimated directly from their 
present expansion, we apply more reliable expansion velocities derived from their
observed sizes. Based on the age of the nebula and the present period of the pulsar, 
we estimate the initial period of the pulsar. If the age of the nebula is unknown, 
we apply the value of 15 ms.
Note, that the final result of our model is not very sensitive on the initial period of 
the pulsar in the case of older nebulae. 
Having fixed these important parameters we can calculate the synchrotron 
spectra produced by leptons inside specific nebulae as a function of their ages. 

From comparison of the calculated synchrotron spectra with the observed  
emission from the Crab Nebula, we estimate the acceleration efficiency of 
leptons by the nuclei and calculate the expected $\gamma$-ray spectra. It is found 
that the considered model can explain the main features of the $\gamma$-ray emission 
from this nebula (see Fig.~1). 
Better description of the Crab Nebula multiwavelength spectrum specifically at 
its lower synchrotron and IC part (in respect to the fitting presented
in Fig.~6 in Bednarek \& Bartosik~2003) is due to the more careful selection of the 
initial parameters of the pulsar (its initial period has been changed to 10 ms). 
The knowledge obtained from modelling of the known $\gamma$-ray
nebulae allows us to predict the level of $\gamma$-ray emission from other nebulae
observed in the radio and X-rays. These low energy observations permit us to 
constrain the efficiencies of lepton acceleration inside these nebulae.

\begin{figure}
  \vspace{7.5truecm}
\includegraphics{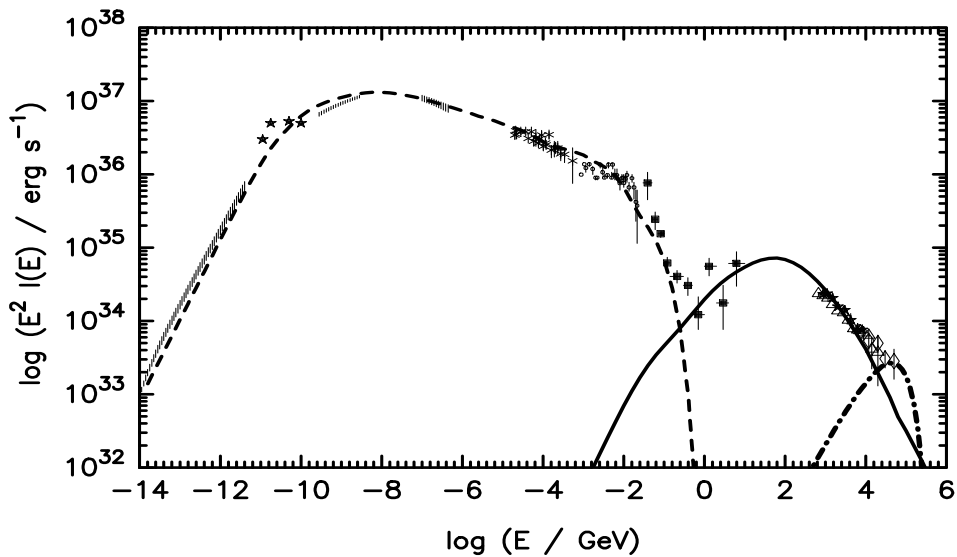}
  \caption{The multiwavelength spectrum of the Crab Nebula compared to the 
spectrum calculated in the hadronic-leptonic model. 
Leptons are injected into the nebula with the power law spectrum (positrons 
accelerated in the interactions with iron nuclei) and spectral index 
$\alpha = 2.5$ between $E_1$ and $E_2$. It is assumed that the iron nuclei 
take $\chi = 0.95$ of the total rotational energy lost by the pulsar. A part of 
the energy of the iron nuclei, $\xi = 0.5$, is converted into relativistic positrons. 
The initial pulsar period is 20 ms. The synchrotron and
inverse Compton spectra (comptonization of synchrotron, MBR, and infrared photons),
produced by leptons inside the nebula, are shown by the dashed and full curves, 
respectively. The $\gamma$-ray spectrum from decay of pions, produced by hadrons 
inside the nebula, is shown by the dot-dashed curve.}
\label{fig1}
\end{figure}

In this paper we consider the population of young pulsars which can be  
observed by Cherenkov telescopes at zenith angles smaller than 
$50^{\rm o}$. This cut-off (e.g. corresponding to the declination of the source 
greater than about $-22^{\rm o}$ for the location of the MAGIC telescope) has been 
introduced in 
order to allow the observations of PWNe with low energy threshold. Our purpose is to 
predict which of them should be the most promissing $\gamma$-ray sources for
the future observations by the next generation of Cherenkov telescopes such as the 
MAGIC or VERITAS. Based on different pulsar catalogs (Taylor et al.~1993, 
Morris et al.~2002, Kramer et al.~2003) and the paper by McLaughlin et al.~(2002,
in the case of PSR J1740+1000) we have selected the classical radio 
pulsars with short periods (below $\sim 200$ ms), strong surface magnetic field
($> 10^{11}$ G), and characteristic ages not significantly larger than $10^5$ yrs.
The parameters of some considered pulsars and their PWNe are shown in Table 1. Note, 
that in the case of some pulsars the surrounding nebulae have not been discovered 
yet. The older pulsars shown in Table 1 has significantly moved from their birth 
place inside the supernova remnant. However this displacement does not have a big 
effect on the estimated here TeV $\gamma$-ray fluxes which are mainly produced 
in older PWNe by leptons scattering the MBR. 

Based on the steady soft X-ray fluxes reported from the nebulae 
surrounding these pulsars ($L_{\rm X,PWN}$) and our modelling of synchrotron emission,
we estimate the efficiency of lepton acceleration by nuclei, $\xi$, in the case of 
every source and calculate the expected $\gamma$-ray fluxes at 
energies above 60 GeV and 200 GeV (see Table 2). The parameter $\chi = 0.95$ of 
energy conversion from the pulsar wind to the relativistic nuclei is kept constant for all 
PWNe.
This procedure is not possible in the case of pulsars with unknown X-ray nebulae 
(B1913+10, J1837-0604, J12740+1000, B1809-19, J1800-21, and J1828-1101). 
Therefore, for these four pulsars (lack of references in Table 1), we apply the 
acceleration efficiencies of leptons, $\xi = 0.15$, derived for the pulsars with the 
closest parameters (i.e B1951+32 for B1913+10 and B1823-13 for five others).


\begin{sidewaystable}
\renewcommand\tabcolsep{.05pc} 
\renewcommand{\arraystretch}{.5} 
\caption{Pulsars and their PWNe}
\begin{center}
\begin{tabular}{|c|c|c|c|c|c|c|c|c|c|} 
\hline
pulsar  & P     & B             & Age   & ${\dot E}$              & distance & PWN/SNR  
& $L_{\rm X,PWN}$(range) & Ref.  & $\xi$ \\
        & (ms)  & $(10^{12}$ G) & (kyr) & ($10^{36}$ ergs/s)& (kpc)    &     
& ($10^{34}$ ergs/s) &  $L_{\rm X}$ &  \\
\hline 
Crab & ~33.4& 3.8          & ~~1.05      &     440  & 
2.0   & Crab Nebula  &  3700~(0.1-4.5keV) & 1  & 0.5 \\
\hline 
J2021+3651& 104. & 3.2        & ~17.      &   ~~3.4    & 
1.5   & G75.2+0.1  & 0.05~(0.3-10keV) & 2 & 0.25 \\
\hline 
J2229+6114& ~51.6& 2.0          & ~10.5      &     ~22.0  & 
3.0   & G106.6+2.9 & 0.1~(0.5-10keV) & 3 & 0.15 \\
\hline
B1951+32& ~39.5   & ~0.49         & 107       &     ~~3.7  & 
2.5   & CTB 80  & 0.27~(0.2-10keV)  &  4 & 0.15 \\
\hline
J0205+6449& ~65. & 3.6        & ~~5.0       &     ~27.0  & 
3.2   & 3C58  &  1.0~(0.5-10keV)  &  5 & 0.1  \\
\hline
B1823-13& 101.5  & 2.8          & ~21        &     ~~2.9   &  
4.1   & G18.0-0.7  & 0.3~(0.5-10keV)  & 6 & 0.15 \\   	    
\hline
B1809-19& ~82.7  & 1.8          & ~51        &     ~~1.8    &  
3.7   & ---  & ---  & --- & 0.15 \\   	    
\hline
B1913+10& ~35.9   & ~0.35         & 170       &     ~~2.9    & 
4.5   & ---  &  ---  & --- & 0.15  \\
\hline 
J1837-0604& ~96.3 & ~2.1        & ~34       &     ~~6.0     & 
6.2   & --- &  ---  & --- & 0.15 \\
\hline 
J1740+1000& 154. &  1.8        & 114       &     ~~0.23  & 
1.4   & --- &  ---  &  --- & 0.15 \\
\hline 
J1800-21& 133.6  & 4.3          & ~16        &     ~~2.2   &  
3.9   & ---  & ---  & --- & 0.15 \\   	    
\hline
J1930+1852& 136. & 10.1        & ~~2.9       &     ~11.8   & 
5.0   & G54.1.+0.3 & 2.1~(2-10keV) & 7 & 0.15 \\
\hline 
J1811-1926& ~65. & ~2.0        & ~2       &     ~~6.4  & 
5.0   & G11.2-0.3  & 1.2~(1-10keV)  &  8 & 0.15 \\
\hline 
J1828-1101& ~72.1 & ~1.05        & ~77       &     ~~1.6  & 
7.2   & --- &  ---  & --- & 0.15 \\
\hline 
\hline
Vela  & ~88.9  & 3.1          & ~11.3        &       ~~6.7   &  
0.3   &  Vela SNR & 0.1~(0.1-4keV)  & 9 & 0.07  \\   	    
\hline
B1706-44& 102.0  & 3.1          & ~17.4      &       ~~3.4   &  
1.8   &  G 343.1-2.3 &  0.03~(0.5-8keV) & 10 & 0.03 \\   	    
\hline 
\multicolumn{10}{l}
{

[1]Harnden \& Seward~(1984);
[2]Hessel et al.~(2004); 
[3]Halpern et al.~(2001);
[4]Li et al.~(2005);
}\\
\multicolumn{10}{l}
{ 
[5]Torii et al.~(2002); 
[6]Gaensler et al.~(2003);
[7]Camilo et al.~(2002);
}\\
\multicolumn{10}{l}
{
[8]Roberts et al.~(2003); 
[9]Becker et al.~(1982);
[10]Romani et al.~(2005);
}\\
\end{tabular} 
\end{center}
\label{tab4}
\end{sidewaystable}

The $\gamma$-ray fluxes from specific PWNe are calculated assuming their
production only in the ICS process by leptons. We neglect possible contribution to 
the $\gamma$-ray spectrum from decay of pions produced in collisions of nuclei 
accelerated by the pulsar with the matter of the supernova envelope. This emission 
is mainly limited to energies above $\sim 10$ TeV and does not have a strong effect in 
the energy range just above $\sim 200$ GeV. 

We consider two models for the soft radiation field inside the nebulae. 
In the first one, leptons scatter only synchrotron radiation (SYN)  
(produced by the same 
population of leptons inside the nebula) and the MBR. 
In the second model, we add additional infrared (IR) component 
created by the warm gas inside the nebula. Such infrared emission with the 
characteristic temperature of $\sim 100$ K is clearly observed in the spectrum 
of the 
Crab Nebula. The infrared target with temperature of $\sim 20$ K is required 
in the case of the nebula around PSR 1706-44 in order to explain the level of 
high energy $\gamma$-ray emission from this object reported by the early observations
(Kifune et al.~1995, Chadwick et al.~1998). However, note that more recent 
observations of PSR 1706-44 by the HESS telescope (Khelifi et al.~2004) does not 
confirm the level of emission reported earlier, what may suggest that additional
infrared target inside PSR 1706-44 is not needed.
The nebulae much older than the nebula around PSR 1706-44 ($> 2\times 10^4$ yrs) 
certainly do not require additional infrared target since their  
infrared emission by warm dust should be relatively weak.

\begin{table}
\begin{center}
\caption{Predicted $\gamma$-ray fluxes above 60 (200) GeV from the PWNe} 
\begin{tabular}{|c|c|c|}
\hline
pulsar   & flux ($10^{-11}{{phot.}\over{cm^2s}}$)  & flux ($10^{-11}{{phot.}\over{cm^2s}}$)    \\
         & (SYN+MBR)           & (SYN+MBR+IR)         \\
\hline 
Crab      &  ---                &     140. (25)             \\
\hline 
J2021+3651&    ~~6 (2)                  &  ~30  (7)     \\
\hline 
J2229+6114&    ~~4 (1)           &     ~16 (4)             \\
\hline
B1951+32 &     ~~9 (1)           & ---                  \\
\hline
J0205+6449&     ~~0.7 (0.2)         &     ~~3 (0.8)             \\
\hline
B1823-13 &     ~~0.7 (0.2)          &     ~~3 (0.6)             \\   	    
\hline
B1809-19 &     ~~1 (0.2)          &     ---           \\   	    
\hline
B1913+10 &     ~~0.9 (0.1)          & ---                   \\
\hline 
J1837-0604&    ~~0.5 (0.1)             &  ~~2 (0.3)    \\
\hline 
J1740+1000&     ~~0.7 (0.08)          &      ---             \\
\hline 
J1800-21  &    ~~0.3 (0.07)            &  ~~1 (0.2)        \\
\hline 
J1930+1852&     ~~0.04 (0.01)            &      ~~0.2 (0.04)            \\
\hline 
J1811-1926&    ~~0.04 (0.02)             &  ~~0.1 (0.04)    \\
\hline 
J1828-1101&    ~~0.07 (0.01)             &  ---     \\
\hline 
\hline
Vela     &     ~37 (9)          &     18 (4)              \\   	    
\hline 
B1706-44 &     ~~0.5 (0.1)            &     ---              \\   	    
\hline 
\end{tabular} 
\label{tab3} 
\end{center}
\end{table} 

In Table 2 we report the PWNe which can be observed by the MAGIC and VERITAS 
telescopes with the highest $\gamma$-ray fluxes at energies above 60 and 200 GeV. 
These threshold values have been selected since they are the approximate lower limits 
of the past Cherenkov telescopes (e.g. Whipple) and the next generation of telescopes 
(e.g. MAGIC, VERITAS). The sources at the top have 
the highest chance of being detected according to our model.
For comparison we also show the results of calculations for the two southern
nebulae, around the Vela pulsar and PSR 1706-44, reported by early observations in 
TeV $\gamma$-rays.
The presence of an additional infrared targets inside the young nebulae can change
significantly, by a factor of $\sim(3-4$), the predicted $\gamma$-ray flux at energies 
below a few TeV (compare the second and third columns in the Table 2). 
The $\gamma$-ray fluxes above 60 GeV are typically a factor of $\sim(4-5)$ higher than
those above 200 GeV.

Only two PWNe on the northern hemisphere (accept Crab) can produce $\gamma$-ray 
fluxes on the level above $\sim 0.1$ of the Crab Nebula flux
(assuming the presence of thermal infrared background as applied for the 
nebula around PSR 1706-44). The second brightest source on the list, 
the pulsar J2021+3651 inside 
the nebula G75.2+0.1, should shine on the level of  $\sim(7-29)\%$ of the Crab Nebula
above 200 GeV, i.e it should be inside the sensitivity limits of the HEGRA and 
the Whipple telescopes. However, note that in the case of this source we have applied 
the distance of 1.5 kpc, estimated from the dimensions of the nebula 
(Hessels et al.~2004), 
but not 12.4 kpc as obtained from the dispersion measure of the pulsar signal 
(Roberts et al.~2002).

There is a tendency, mentioned and discussed already by Aharonian et al.~1997, that 
pulsars with 
weaker surface magnetic fields, and hence lower magnetic fields inside their nebulae, 
are relatively stronger high energy $\gamma$-ray emitters. For example, compare the Crab 
pulsar with the PSR B1951+32. They have similar periods and distances, but differ by an 
order of magnitude in their surface magnetic field. The rotational energy loss of 
PSR B1951+32 is two orders of magnitude lower but the $\gamma$-ray flux is only $\sim$20 
times lower in spite of lower, by a factor of 3, efficiency of lepton acceleration.
The same feature is clearly seen in the case of J1811-1926 and J0205+6449.
For the same reason, the second youngest considered pulsar, J1930+1852, with the 
characteristic age of 2900 yrs, is also relatively inefficient $\gamma$-ray source. 
This is due to the fact that the high energy leptons injected in the past 
cool efficiently in a strong nebular magnetic field. They can not accumulate inside the 
nebula for a long time and that's why can not produce strong high energy $\gamma$-ray 
fluxes at the present time.  

\section{Discussion and Conclusion}

We predicted top 10 pulsar wind nebulae which should be observable in TeV $\gamma$-rays
at the Northern 
hemisphere with the highest fluxes at energies above 60 GeV and 200 GeV (Table 2).
None of the considered PWNe turns out to be comparable in brightness to the Crab 
Nebula. 
If the effective collection area of the constructed telescopes (MAGIC, VERITAS) is
$\sim 10^5$ m$^2$ down to 60 GeV then the 10 top PWNe should produce $\gamma$-ray rates 
above $\sim 25\gamma$/hr at energies above 60 GeV. However, the actual detection 
capabilities 
will differ for specific experiments since they depend on the efficiency of  
$\gamma$-ray selection criteria from the hadronic background.  

Present observations do not constrain the predictions of our model for the pulsars
observable from the Northern hemisphere (listed in Table 1). The upper limits 
available in the literature (although at higher energies) are above fluxes
predicted in this paper, see e.g. 
PSR 1951+32 (Srinivasan et al.~1997), PSR J0205+6449 (Hall et al.~2001),  
PSR B1823-13 (Hall et al. 2003, Fegan et al.~2005, Aharonian et al.~2005c), 
or  PSR 2229+6144, PSR J2021+3651, PSR B1823-13 (Fegan et al.~2005). 
However, note that some of these upper limits are quite close to the fluxes predicted 
in the model with additional infrared background inside the nebula (e.g.
for J2229+6144 and J2021+3651, Fegan et al.~2005 and PSR B1823-13, 
Aharonian et al.~2005). Recently the HESS group reports the 
upper limits for the nebulae around the two southern pulsars, Vela and PSR 1706-44 
(Khelifi et al 2005, Aharonian et al.~2005) on the level of only a few percent of 
the Crab Nebula flux, i.e.  significantly lower than the earlier positive reports 
(Yoshikoshi et al. 1997, Kifune et al. 1995, 
Chadwick et al. 1998). If confirmed, the model with an additional infrared target
inside the nebula around PSR 1706-44 would predict the TeV $\gamma$-ray 
fluxes too high (see Bednarek \& Bartosik~2003).
In the case of the Vela pulsar the confrontation of our model, predicting the TeV 
flux on the level of $\sim 30\%$ of the Crab Nebula, with the new 
observations is more complicated due to the proximity of the Vela pulsar to the Sun. 
Our modeling shows that the magnetic field in the main volume of the Vela nebula is 
comparable to the magnetic field inside the interstellar medium (except for only a 
very small region around the pulsar). We estimate the dimension of the 
TeV source in the case of the Vela nebula by calculating the diffusion distance of 
leptons with energies 100 TeV during the age of the nebula ($\sim 10^4$ yrs) in the 
magnetic field of $\sim 3\times 10^{-6}$ G. The Bohm diffusion coefficient has been 
applied. It is found that leptons can diffuse as far as $\sim 5$ pc from the pulsar. 
However 
the synchrotron and IC energy loss time scales for 100 TeV leptons in this magnetic 
field and the MBR are comparable to the age of the Vela nebula. Therefore,
leptons with energies 100 TeV can diffuse from the pulsar up to $\sim 5$ pc
without drastic energy losses.
For the distance of the pulsar from the Sun equal to $\sim 300$ pc, the 
TeV $\gamma$-ray source with dimension $\sim 5$ pc should have the angular extend 
on the sky 
as large as $\sim 1^{\rm o}$. Therefore, we conclude that the TeV source around 
the Vela pulsar should be quite extended on the sky and 
that's why difficult to observe by the Cherenkov telescopes. 

When estimating the TeV $\gamma$-ray fluxes from the population of the PWNe,
we applied the basic parameters of their supernova remnants, e.g. the expansion 
velocity and the mass of exploding supernova, as derived for the Crab Nebula.
In fact, these parameters are mainly unknown for these nebulae. 
The increase of the initial mass of the expanding supernova has an effect on the 
$\gamma$-ray flux produced by relativistic nuclei in collisions with the matter
at energies above a few TeV but gives negligible effect on the fluxes at 
$\sim 200$ GeV. The increase of the initial expansion velocity effects the dimension 
of the nebula at a specific time after explosion.
Larger nebula means weaker magnetic field and, as we discussed above, larger fluxes
of TeV $\gamma$-rays due to the less efficient synchrotron cooling of relativistic 
leptons.

\section*{Acknowledgments}
This work is supported by the Polish MNiI grants No. 1P03D01028 and 
PBZ KBN 054/P03/2001.

\end{document}